\documentclass[twocolumn, showkeys,amsmath,amssymb]{revtex4}

\newcommand{\qmass}{\ensuremath{\frac{\hbar}{c} \sqrt{\frac{\Lambda}{3}}}}

\begin{document}

\title{Holography and non-locality in a closed vacuum-dominated universe}

\author{T. R. Mongan}
\affiliation{84 Marin Avenue \\
			 Sausalito, CA 94965 USA}
\email{tmongan@mail.com}


\begin{abstract}
A closed vacuum-dominated Friedmann universe is asymptotic to a de Sitter space
with a cosmological event horizon for any observer.  The holographic principle
says the area of the horizon in Planck units determines the maximum number of
bits of information about the universe that will ever be available to any
observer.  The wavefunction describing the probability distribution of mass
quanta associated with bits of information on the horizon is the boundary
condition for the wavefunction specifying the probability distribution of mass
quanta throughout the universe.  Local interactions between mass quanta in the
universe cause quantum transitions in the wavefunction specifying the
distribution of mass throughout the universe, with instantaneous non-local
effects throughout the universe.
\end{abstract} 


\maketitle

It now seems inescapable that quantum mechanics is fundamentally non-local.
However, a mechanism for non-locality (Einstein's ``spooky action at a
distance'') has been lacking.  This paper suggests that the holographic
principle indicates a possible mechanism for non-locality in any closed
vacuum-dominated Friedmann universe.  Specifically, a holographic non-local
quantum-mechanical description can be developed for the finite amount of
information in a closed vacuum-dominated universe.  It is assumed the universe
began by a quantum fluctuation from nothing, underwent inflation and became so
large that it is locally almost flat.  It is also assumed that, after inflation,
the vacuum energy density of the universe is constant in space and time (i.e.,
there is a cosmological constant).  One way such a universe can arise is
outlined in the quantum cosmology presented in Ref.~\cite{mongan:2001}.  When it
began, the closed universe contained all the information it will ever contain.
There is nothing outside a closed universe, so no information can come into the
universe from elsewhere.

At late times in a vacuum-dominated universe, the Friedmann equation becomes, 
\[
	\left(\frac{\dot R}{R}\right)^2 
	= \frac{8 \pi G \varepsilon_\nu}{3c^2} 
	= \frac{\Lambda c^2}{3}, 
\]
where the cosmological constant $\Lambda$ is related to the vacuum
energy density by
\[ 
	\Lambda = \frac{8\pi G \varepsilon_\nu}{c^4},
\]
and the universe is asymptotic to a flat de Sitter space.  There is a
cosmological event horizon in de Sitter space, at a radial distance $R_F = c/H =
\sqrt{3/\Lambda}$ from any observer (see, e.g.,
Refs~\cite{carniero:2003},~\cite{padmanabhan:2002} and~\cite{frolov:2003}).
Taking the Hubble constant as $H_0 = \mathrm{65 \, km \, sec^{-1} \, Mpc^{-1}}$,
the critical density $\rho_c = \frac{3H_0^2}{8\pi G} = \mathrm{7.9 x 10^{-30} \,
g \, cm^{-3}}$, the vacuum energy density $\varepsilon_\nu = 0.7 \rho_c c^2 =
\mathrm{5.0 x 10^{-9} \, g \, cm^{2} \, sec^{-2} \, cm^{-3}}$, and $\Lambda =
\mathrm{1.0 x 10^{-56} \, cm^{-2}}$.  Then the de Sitter horizon radius is $R_F
= \mathrm{1.7 x 10^{28} cm}$ and the area of any observer's de Sitter horizon is
$A = 4\pi R_F^2 = 12\pi /\Lambda$.

Black hole thermodynamics led to the holographic principle~\cite{bousso:2002},
indicating that the number of bits of information inside a horizon with surface
area $A$ is $N = A/(4\ln 2)$, where $A$ is measured in Planck units.  Applied to
the de Sitter horizon, the holographic principle indicates the total number of
bits of information that will ever be available to any observer in the universe
is
\[
	N_F = \frac{A}{4\ln 2} 
	= \frac{\pi R_F^2}{\ln 2} 
	= \frac{3\pi}{\Lambda \ln 2}
\]
in Planck units.  So, 
\[
	N_F = \frac{\pi R_F^2}{\delta^2 \ln 2} 
	= \frac{3\pi}{\Lambda \delta^2 \ln 2} 
	= \mathrm{5.2 x 10^{122}}, 
\]
where the Planck length $\delta = \sqrt{\frac{\hbar G}{c^3}} = \mathrm{1.62 x
10^{-33} \, cm}$, and $N_F$ is also the number of pixels of area $4\delta^2 \ln
2$ on any observer's de Sitter horizon.

Reasonable physical theories must be based only on information available to an
observer, and the de Sitter horizon specifies the total amount of information
that will ever be available to any observer.  Each observer has a different de
Sitter horizon, but (consistent with relativity) no observer has a privileged
position.  When we are eventually able (in principle) to observe the de Sitter
horizon, the holographic information on the horizon will characterize the state
of the universe at the instant, billions of years previously, when the photons
carrying the information left the horizon.

After the first few seconds of the life of the universe, the energy exchanged
between matter and radiation was negligible compared to the total energy of
matter and radiation separately~\cite[page 726]{misner:1973}.  In a closed
universe, when the exchange of energy between matter and radiation is
negligible, the total mass of the universe and the total number of massless
quanta in the universe are conserved~\cite{misner:1973}.  There are about
$\mathrm{10^{90}}$ massless quanta in a closed universe with today's radiation
energy density, a radius of about $\mathrm{10^{28}}$ cm and a cosmic microwave
background temperature of 2.73 \textsuperscript{o}K.  This number is negligible
compared to the total number of degrees of freedom in the universe, so the
following analysis focuses on a quantum mechanical description of the
distribution of the mass within the universe and neglects massless degrees of
freedom.

Describing a universe with only a finite number of degrees of freedom is an
extremely difficult problem for quantum field theory, so this analysis is not
based on quantum field theory.   A quantum mechanical description of the finite
number of bits of information on a horizon requires wavefunctions specifying the
probability distribution of those bits of information on the horizon.  This
analysis considers a description of the universe that can in principle be
developed by observers using the finite amount of information available on their
de Sitter horizon.

Wesson~\cite{wesson:2004}  notes that two mass scales, a quantum mass scale and
a gravitational mass scale, can be formed from the constants $G, \hbar, \Lambda$
and $c$.  The quantum mass scale $\qmass = \mathrm{2.0 x 10^{-66} g}$ is
the scale for the minimum quantum of mass in the universe.  The gravitational
mass scale $\frac{c^2}{G} \sqrt{\frac{3}{\Lambda}} = \mathrm{2.3 x 10^{56} g}$
is the scale for the total conserved mass of a closed universe. The holographic
principle says there are $3\pi /(\Lambda \delta^2 \ln 2)$ observable bits of
information on the de Sitter horizon at radius $\sqrt{3/\Lambda}$ in a universe
dominated by a cosmological constant $\Lambda$.  If the total conserved mass of
the closed universe is $b\frac{c^2}{G} \sqrt{\frac{3}{\Lambda}}$, and there are
$3\pi /(\Lambda \delta^2 \ln 2)$ bits of information on the de Sitter horizon,
the mass associated with each bit is 
\[
    b \frac{c^2}{G} \sqrt{\frac{3}{\Lambda}} / 
	\frac{3\pi}{\Lambda \delta^2 \ln 2}  
	= 
	\frac{b\ln 2}{\pi} \qmass.
\]
So, the holographic principle indicates that the mass scale for a bit of
information is the quantum mass scale $\qmass = \mathrm{2.0 x 10^{-66} g}$ for
the minimum quantum of mass in the universe.  The quantum states with the
longest wavelength $2\pi R_F = 2\pi \sqrt{3/\Lambda}$ on the de Sitter
horizon correspond to states with the lowest energy on the horizon and should be
associated with the quantum of mass.  If the quantum of mass is  $\qmass$ that
mass quantum has Compton wavelength $2\pi R_F$.  This suggests that $\frac{b\ln
2}{\pi} = 1$ and each bit of information is associated with a quantum of mass
$\qmass$ with Compton wavelength $2\pi R_F$.

As noted previously, the number of bits of information in the universe  $N_F =
\frac{4\pi R_F^2}{4\delta^2 \ln 2} = \mathrm{5.2 x 10^{122}}$ is also the number
of pixels of area $4\delta^2 \ln 2$ on any observer's de Sitter horizon.  So, a
quantum description of the total amount of information available about the
universe can be obtained by identifying each area (pixel) of size $4\delta^2 \ln
2$  on the de Sitter horizon with one bit of information, associated with the
wavefunction for a quantum of mass $\qmass$ with Compton wavelength $2\pi R_F$ .
If the $z$ axis (in spherical coordinates centered on the observer's position)
pierces the center of the $i$th pixel of area $4\pi \delta^2 \ln 2$  on the
horizon with radius $R_F$, $\Theta_i$ is the polar angle measuring the angular
distance from the $z$ axis through the $i$th pixel to the point on the horizon
where the wavefunction is evaluated and $C_2$ is a normalization constant,
wavefunctions of the form $C_2\cos \Theta_i$  on the de Sitter horizon have the
necessary wavelength $2\pi R_F$ and can define the probability distribution for
finding the bit of information associated with the \textit{i}th pixel at any
location on the horizon.  The chosen form of the wavefunction $C_2\cos \Theta_i$
insures that the maximum probability of finding the \textit{i}th bit of
information on the de Sitter horizon is in the two pixels on opposing
hemispheres where the $z$ axis of that wavefunction pierces the horizon.  These
wavefunctions have anti-nodes where the $z$ axis intercepts the horizon and a
nodal line around the equator (polar angle = $\pi /2$ ).

Regarding the wavefunctions associated with two diametrically opposite pixels,
if +1 is assigned to pixels containing a ``peak'' anti-node and -1 is assigned
to pixels containing a ``valley'' anti-node, two situations are possible.  If
two pixels diametrically opposite each other on different hemispheres of the
horizon have the same sign, the probability waves destructively interfere all
over the surface, resulting in zero probability of finding mass quanta
associated with a bit of information at any location on the horizon.  If two
diametrically-opposed pixels have opposite signs, the wavefunctions
constructively interfere, resulting in a wavefunction $\psi_i =
\sqrt{\frac{3}{2\pi R^2_F}}\cos \Theta_i$  normalized to two mass quanta on the
horizon.  In this way, the constructive and destructive interference of the
$\cos \Theta_i$ wavefunctions on the horizon requires that mass quanta occur in
pairs.  So no additional bits of information are required to encode the
information in the universe if each member of a pair of mass quanta has the
opposite value of a single (conserved) quantum number.

The quantum of mass $\qmass$ is far below the mass scales familiar from
experimental particle physics.  The proton mass, $m_p = \mathrm{1.67 x 10^{-24}
\, g = 938 \, MeV}$ is about $\mathrm{10^{42}}$ times the quantum of mass and
the electron mass 0.511 Mev  is about $\mathrm{5 x 10^{38}}$ times the quantum
of mass.  A mass as small as 1 eV is about 10\textsuperscript{33} times the
quantum of mass, and the current upper limit on the mass of the photon is about
fifteen orders of magnitude larger than the mass quantum.  So, there are at
present no theories relating the quantum of mass to the familiar masses observed
in particle physics experiments.  However, if each member of a pair of mass
quanta has opposite values of a single as yet unknown quantum number, composite
massive systems described by a superposition of wavefunctions for pairs of mass
quanta need not be identical except for their mass.  A theory involving mass
quantum pairs with opposite values of a single quantum number may allow
properties like spin and charge to emerge as quantum numbers characterizing
composite quantum mechanical systems involving the large numbers of mass quanta
necessary to make the ``elementary'' particles we have observed to date.

The distribution of mass quanta associated with bits of information on the
horizon at any instant of cosmic time $t$ can be represented by a
two-dimensional lattice gas involving $\frac{N_F}{2}$ mass quantum pairs where
each lattice gas site has area $4\delta^2 \ln 2$.  The lattice gas has zeros in
half of the pixels on the horizon and ones in the remaining pixels.  Pixels
diametrically opposite to each other on the horizon are constrained to have the
same value.  Zeros correspond to situations where wavefunctions from
diametrically-opposed pixels on the entire horizon destructively interfere to
produce zero probability of finding mass quanta associated with bits of
information related to those pixels at any location on the horizon.  The
remaining pixels, with a value 1, contain an anti-node of the wavefunction
$\psi_i = \sqrt{\frac{3}{2\pi R^2_F}}\cos \Theta_i$ normalized to two mass
quanta on the horizon of radius $R_F$.  

The probable number $n$ of the mass quanta associated with one bit of
information in an area $A$ on the horizon is 
\[
	n(A) = N_F\int \!\! \int_{A}{}
		\left|\prod_{i=1}^{N_F} a_i \psi_i\right|^2 dA, 
\]
based on the lattice gas representation of the information on the horizon. This
is also the probability of finding $n(A)$ of the mass quanta associated with a
bit of information in the solid angle within the universe that subtends the area
$A$ on the horizon and has its apex at the observer's position. The $N_F$
numbers $a_i$ associated with each pixel on the horizon are either zero or one,
the numbers corresponding to pixels diametrically opposite to each other on the
horizon are constrained to be equal, and together these numbers encode all the
information that will ever be available about the universe.

The wavefunctions on the horizon are the boundary condition on the form of the
wavefunctions specifying the probability distributions of the finite number of
mass quanta distributed throughout the featureless background space of a closed
universe with a constant vacuum energy density (cosmological constant).  The
wavefunction for the probability of finding a mass quantum anywhere in the
universe is a solution to the Helmholtz wave equation in the closed universe.  A
closed universe with radius of curvature $R$ can be defined by the three
coordinates $\chi, \Theta, \mathrm{and } \phi$ where the volume of the
three-sphere $(S^3)$ is~\cite{islam}
\[
	R^3 \int_{0}^{\pi}\sin^2 \chi d\chi \int_{0}^{\pi}\sin \theta d\theta
		\int_{0}^{2\pi}d\phi
\]
.  The wavefunction for a bit of information on the horizon is the
projection of the wavefunction for that bit of information in the volume of the
universe on the horizon at $R_F = R\sin \chi$.  One solution to the Helmholz
equation on the three-sphere~\cite{gerlach:1978}  is the scalar spherical
harmonic $Q_{10}^2 = \sqrt{\frac{12}{\pi}}\cos \Theta_i \sin \chi$.  For the
$i$th degree of freedom on the horizon, the $\cos \Theta_i$ behavior of the
wavefunction on the horizon determines the wavefunction $\psi_i = C_3
Q_{10}^2(i) = C_3\sqrt{\frac{12}{\pi}}\cos \Theta_i \sin \chi$ as the solution
of the Helmholz wave equation describing the probability distribution within the
universe of the mass quanta associated with the $i$th bit of information.   The
constant $C_3 = \sqrt{\frac{4}{\pi^2 R^3}}$ is determined by normalizing the
wavefunction to two quanta of mass within the universe.  The probable number $n$
of the mass quanta associated with one bit of information in any volume $V$
within the universe is
\[
	n(V) = N_F \int \! \int \! \int_V 
		\left| \prod_{i=1}^{N_F} a_i \psi_i \right|^2 dV.
\]
So, the information specified by the finite quantum lattice gas representation
on the horizon determines the probability of finding a certain number of mass
quanta in any given volume within the universe.  

Quantum state changes caused by local interactions between mass quanta have
non-local consequences throughout the universe.  Any local change in the quantum
state of the mass distribution within the universe is instantly reflected in
changes in the eigensolutions of the Helmholtz wave equation within the
universe, as well as in the lattice gas representation of the universe on an
observer's de Sitter horizon.  This instantly changes both the probability
distribution of the bits of information on the horizon and the corresponding
probability distribution of mass quanta throughout the universe.  In this way,
the holographic principle indicates a mechanism for non-locality in quantum
processes throughout the universe.

If the universe is closed, it is now so large that it is locally flat.  At
present, we can't see beyond the surface of last scattering that displays
information from the era when photons decoupled from matter and began streaming
freely through the universe.  However, as we look out into the universe,
Bousso's light sheet analysis~\cite{bousso:2002} indicates that data from
observations at redshift $z$, corresponding to a proper distance $D$ from us, can
be used to develop a holographic description of the universe at the
corresponding lookback time.  Because the area of the surface at distance $D$ is
smaller than the area of the de Sitter horizon, the description will be more
coarse grained than the description based on information from the deSitter
horizon.  The light sheet analysis indicates that the number of bits of
information available to any observer on the spherical surface at distance $D$
from us is 
\[
	N_E = \frac{\pi D^2}{\delta^2 \ln 2} = \frac{\pi c^3 D^2}{\hbar G \ln 2}.
\]
The mass associated with each bit of information is then, 
\[
	\left(\frac{\pi c^2}{G\ln 2}\sqrt{\frac{3}{\Lambda}}\right) 
	\left(\frac{\hbar G\ln 2}{\pi c^3 D^2}\right) 
	= \frac{\hbar}{c}\left(\frac{\sqrt{3/\Lambda}}{D^2}\right),
\]
and this is larger than the quantum of mass $\qmass$ until $D = R_F =
\sqrt{3/\Lambda}$.   The holographic principle again indicates the probability
distribution of mass within the universe is encoded by a lattice gas
representation involving $N_E$ sites on the spherical surface at distance $D$
from the observer.  The number of bits of information available for describing
the probability distribution of mass within the universe goes up and the average
mass associated with those bits of information goes down as the distance $D$
increases, until the de Sitter horizon is reached.

This analysis may help to address the quantum measurement problem.  The
holographic principle suggests a very close linkage between the information
available about a physical system and the system itself.  The analysis provides
a quantum mechanical description of the information available to describe
physical systems in a closed vacuum-dominated universe.  Greenstein and
Zajonc~\cite{greenstein:1997} then suggest the measurement problem is
considerably eased if quantum mechanics is viewed as a theory describing the
evolution of the \emph{information} available to describe physical systems.


\begin{thebibliography}{99}

\bibitem{mongan:2001} T. Mongan, Gen. Rev. Grav. \textbf{33}, 1415-1424
(2001), [gr-qc/010302].

\bibitem{carniero:2003} S. Carniero, Grav. Res. Found. (2003), special issue
of the IJMPD, [gr-qc/0305081].

\bibitem{padmanabhan:2002} T. Padmanabhan, Class. Quant. Grav. \textbf{19},
5387-5408 (2002), [gr-qc/0204019].

\bibitem{frolov:2003} A. Frolov and L. Kofman, JCAP \textbf{0305}, 009 (2002),
[hep-th/0212327].

\bibitem{bousso:2002} R. Bousso, Rev. Mod. Phys. \textbf{74}, 825 (2002),
[hep-th/0203101].

\bibitem{misner:1973} C. W. Misner, K. S. Thorne, and J. A. Wheeler,
\emph{Gravitation} (W. H. Freeman and Company, New York, 1973).

\bibitem{wesson:2004} P. Wesson, Mod. Phys. Lett. \textbf{A19}, 1995-2000
(2004), [gr-qc/0309100].

\bibitem{islam} J. Islam, \emph{An Introduction to Mathematical Cosmology}
(Cambridge University Press), 2nd edition, pages 43-44.

\bibitem{gerlach:1978} U. Gerlach and U. Sengupta, Phys. Rev. \textbf{D18},
1773 (1978).

\bibitem{greenstein:1997} G. Greenstein and A. Zajonc, \emph{The Quantum
Challange} (Jones and Bartlett Publishers, Sudbary, Massachusetts, 1997), page
188.

\end{thebibliography}
\end{document}